\documentstyle[epsfig,epsf,psfig]{mn2e}

\title[Electron-cyclotron maser emission from white-dwarf
  systems]{Electron-cyclotron maser emission from white-dwarf pairs
  and white-dwarf planetary systems}

\author[Andrew J. Willes, Kinwah Wu]
  {Andrew J. Willes$^{1}$\thanks{E-mail: willes@physics.usyd.edu.au}  and
  Kinwah Wu$^{2}$\thanks{E-mail: kw@mssl.ucl.ac.uk}  \\ 
$^{1}$School of Physics, University of Sydney, NSW 2006, Australia\\
$^{2}$Mullard Space Science Laboratory, University College London,
      Holmbury St Mary, Dorking, Surrey RH5 6NT}

\begin{document}

\date{ }

\pagerange{\pageref{firstpage}--\pageref{lastpage}} \pubyear{2002}

\maketitle

\label{firstpage}

\begin{abstract}

By analogy to Jovian radio emissions 
  powered by the electromagnetic interaction between Jupiter and its moons, 
  we propose that close magnetic-nonmagnetic white-dwarf pairs 
  and white-dwarf planetary systems 
  are strong radio sources.
A simple model is developed to predict the flux densities of radio emission 
  generated by a loss-cone-driven electron-cyclotron maser.  
The radio emission from these systems 
has high brightness temperatures, 
  is highly polarized, 
  and varies on a periodic cycle following the orbital rotation. 
Masers from magnetic-nonmagnetic white-dwarf pairs, 
  with orbital periods $\la 10$ min, are expected 
  to be detectable over a wide range of radio frequencies.  
Terrestrial planets in close orbits about magnetic white dwarfs, 
  with orbital periods $\la 30$ hr, can also produce 
  detectable radio emission, 
  thus providing a means to identify 
Earth-sized extrasolar planets.
\vspace*{2cm}
\end{abstract}

\begin{keywords} 
   masers -- 
   radiation mechanism: non-thermal -- 
   stars: binaries: close -- 
   stars: white dwarfs -- 
   planetary systems -- 
   radio continuum: stars 
\end{keywords}

\section{Introduction}

The operating requirements for electron-cyclotron masers are
  (i) a population inversion in the electron distribution
  and (ii) a magnetized plasma in which
  the electron-cyclotron frequency $\Omega_{\rm e}$
  exceeds the plasma frequency $\omega_{\rm p}$
  (e.g. Dulk 1985).
The first condition can be achieved
  when the magnetic field geometry in the source region 
  allows the development of an anisotropy in the electron distribution.
One common example is the loss-cone electron distribution
  (e.g.\ Wu \& Lee 1979), 
  where an electron pitch-angle anisotropy develops 
  within a magnetic flux tube with converging field lines at each footpoint.
Large pitch angle electrons are magnetically reflected, 
  whereas small-pitch-angle electrons are lost through collisions 
  with high density plasma at the foot of the flux tube.
The second condition is satisfied
  in magnetized plasmas with a relatively low electron density
  and/or high magnetic field strength.

These conditions are satisfied in a variety of astronomical systems,
  including the solar corona (Melrose \& Dulk 1982; Robinson 1991),
  terrestrial and planetary magnetospheres 
  (Zarka 1992; Farrell, Desch \& Zarka 1999: Ergun et al.\ 2000),
  and solar-like stars and magnetically-active binary stars
  (e.g.\ Melrose \& Dulk 1982; Abada-Simon et al.\ 1994;
  Slee, Haynes \& Wright 1994; Trigilio et al.\ 1998).
Electron-cyclotron maser emission is characterized 
  by  high degrees of circular polarization (typically 
nearly $100\%$), 
  high brightness temperatures ($T \gg 10^{8}$~K),  
  and narrow beaming. 
  
Planetary radio emissions are typically solar-wind driven, 
  with the exception of a significant proportion of Jovian radio emissions, 
  which are correlated with the orbital phase of Io (Bigg 1964),
  and to a lesser extent with the phases of Ganymede and Callisto 
  (Menietti et al.\ 1998, 2001). 
In the unipolar-inductor model for the Jupiter-Io interaction
  (Piddington \& Drake 1968; Goldreich \& Lynden-Bell 1969)
  a current circuit is set up along the magnetic field lines 
  connecting Io to Jupiter, closing in the Jovian ionosphere. 
The circuit is driven by the e.m.f. induced across Io, 
  which is a conducting body,  
  as it traverses the Jovian magnetic field.
The existence of a Jupiter-Io current circuit was confirmed 
  by the detection at ultraviolet and infrared wavelengths
  of a footprint in Jupiter's upper atmosphere,
  which remains fixed at the position of the footpoint of the Io flux tube 
  as Jupiter rotates (Connerney et al.\ 1993; Clarke et al.\ 1996).
The unipolar-inductor model predicts the existence of point-like auroral footprints. 
The observed tail of ultraviolet emission, extending up to $100^{\circ}$ 
  in longitude downstream of the bright Io footprint (Clarke et al.\ 2002), 
  reveals that plasma-inertia effects are also important 
  in the Jupiter-Io interaction, in addition to the unipolar inductor.
The plasma-inertia effects are due to
  the presence of a dense plasma torus about Io's orbit,
  consisting of ionized volcanic material from Io, 
  which is ``picked up'' into corotation 
  with the rotating Jovian magnetic field (Brown 1994).
The auroral ultraviolet footprints 
  for the Jovian moons Ganymede and Europa
  appear as point-like sources (Clarke et al.\ 2002),
  consistent with the unipolar-inductor model, 
  despite the fact that  the Ganymede-Jupiter interaction
  is complicated by Ganymede's embedded magnetosphere (Kivelson et al.\ 1996). 
An auroral footprint for Callisto has not been detected 
  because the footpoint of the Callisto flux tube 
  is coincident with the main auroral oval (Clarke et al.\ 2002). 

Radio emission is generated in the unipolar-inductor model 
  by electron cyclotron maser emission 
  from current-carrying electrons in the Io flux tube. 
  The free energy for maser emission is provided 
  by a loss-cone instability in the reflected electron velocity distribution. 
The observed anti-correlation 
  between infrared footpoint emission and Io-controlled Jovian decametric radiation 
  supports the hypothesis that the radio emission is driven
  by reflected electrons (Connerney et al.\ 1993).
In addition, the negative frequency drifts evident  
  in the fine-frequency structure of Jovian decametric radiation 
  (the Jovian S-bursts) is consistent with maser emission driven 
  by reflected electrons (Ellis 1974).
Theoretical models based on electron-cyclotron maser emission 
  can explain the observed high brightness temperatures ($\ga 10^{17}$~K, Dulk 1970), 
  $100\%$ elliptical polarization (Dulk, Lecacheux \& Leblanc 1992)  
  and the radiation beaming pattern.
Ground-based and dual spacecraft observations 
  have confirmed the theoretical prediction 
  that Io-controlled emission consists of hollow conical radiation beams, 
  with opening angles $\la 80^{\circ}$ and beam widths $\la 2^{\circ}$, 
  with the source location at the foot of the Io flux tube 
  (Dulk 1967; Hewitt et al.\ 1981; Maeda \& Carr 1992; Kaiser et al.\ 2000).

Recently, two ultra-short-period ($P < 10~{\rm min}$)
  X-ray sources RX~J1914+24 (Cropper et al.\ 1998; Ramsay et al.\ 2001, 2002b)
  and RX~J0806+15 (Israel et al.\ 2002; Ramsay, Hakala \& Cropper 2002a)
  were discovered.
The remarkable characteristics of these sources 
  are that there is only a single period 
  for both the X-ray and optical emission,
  which is interpreted as the orbital period (Cropper et al.\ 1998)
  and the X-ray and optical emission are anti-phased (see Ramsay et al.\ 2001).
The inferred orbital periods are too short 
  to be from a binary system containing a white dwarf and a main-sequence star.
Consequently, these X-ray sources are proposed to be compact binaries
  containing two white-dwarf stars, 
  which may be in synchronous rotation with the orbit (Cropper et al.\ 1998).
Their X-ray and optical/IR variations
  are contrary to those of typical accretion-powered sources,
  which often show multiple periods,
  corresponding to the orbital rotation, spin of the accreting star and beats.
Moreover, the absence of optical polarization and helium emission lines 
  does not support the scenario involving mass transfer and accretion.

Wu et al.\ (2002) proposed a unipolar-inductor model
  to explain the peculiar X-ray and optical/IR variations of RX~J1914+24, 
  based on the unipolar-inductor model
  for the Jupiter-Io system.
In the unipolar-inductor model for white-dwarf pairs, 
  Jupiter is replaced by a magnetic white dwarf,  
  and Io is replaced by a nonmagnetic white dwarf. 
The X-ray emission is powered by
  resistive heating of the magnetic white-dwarf atmosphere
  at the footpoints of the magnetic flux-tubes
  carrying the electric currents flowing between the white dwarfs.
This is analogous to the ultraviolet footpoint emission of the Galilean moons.   
The anti-phased optical/infrared emission emanates 
  from the irradiated face of the nonmagnetic white dwarf, 
  which is visible when the X-ray source is directed away from the observer.  
The orbital period gradually decreases in the unipolar-inductor model 
  because the orbital evolution is driven by power losses 
  through gravitational radiation, 
  and is therefore consistent with timing measurements 
  of the X-ray pulses from RX J1914+24, 
  in which the orbital period decreases at a rate 
  consistent with gravitational radiation losses (Strohmayer 2002). 
This is in contrast to models 
  for the X-ray emission involving mass transfer from one white dwarf to another, 
  which predict an increase in the orbital period.  
In contrast to the Jupiter-Io system,
  plasma-inertia effects are unlikely to be significant in the white-dwarf context 
  because of the relatively high magnetic field strengths and low plasma densities
  (Wu et al.\ 2002).
  
It has also been proposed that unipolar induction 
  can operate in isolated magnetic white-dwarf systems with terrestrial planets 
  (Li, Ferrario \& Wickramasinghe 1998).   
In this case the planet plays the role of the non-magnetic white dwarf 
  in magnetic-nonmagnetic white-dwarf binary systems. 
The induced e.m.f. and dissipated power 
  in the white-dwarf systems is significantly higher than in the planet-moon systems 
  because the magnetic moment is higher for white dwarfs than for planets,
  and both white dwarfs and planets are typically larger than planetary moons. 

We predict here, by direct analogy with the Jupiter-moon systems,  
  that compact magnetic-nonmagnetic white-dwarf pairs
  and planets orbiting magnetic white dwarfs  
  are strong electron-cyclotron maser sources. 
The large e.m.f. produced by the orbital motion of the system
  supplies high-energy electrons, 
  and the (dipolar) geometry of the white-dwarf magnetic field
  and the low-plasma-density environment 
  between the two stars or between the star and the planet  
  can develop instabilities that generate electron-cyclotron maser emission.

Around 100 extrasolar planets are currently known. 
With the exception of pulsar planets, 
  these planets,  
  which have masses comparable to Jupiter and orbit around solar-like stars,  
  were detected with the radial-velocity technique. 
Other methods proposed for detecting Jovian (gas-giant) planets 
  around white dwarfs include observation 
  of the excess thermal infrared emission associated with the Jovian planet 
  (Ignace 2001; Burleigh, Clarke \& Hodgkin 2002), 
  or of re-emitted hydrogen recombination lines 
  from the planetary atmosphere for close planetary orbits ($\la$ a few AU)
  around a hot white dwarf (Chu et al. 2001).
We propose here 
  that radio observations can reveal the presence of terrestrial planets 
  in close orbits around white dwarf stars. 
These systems are remnants of main-sequence planetary systems 
  in which terrestrial planets survive the stellar expansion phases 
  and migrate to a stable close orbit around the white dwarf.

This paper is organized as follows.
In \S 2 we use a simple model to calculate flux densities for maser emission,
  assuming parameters appropriate in the environment of white-dwarf pairs
  and of white-dwarf planetary systems.
A more self-consistent consideration
  which takes into account the geometry and properties
  of the current sheets along the flux tubes
  will be presented in a future paper. 
In \S 3 we discuss briefly the astrophysical implications
  of the results obtained from our calculations.

\section{Model outline}

In this Section, 
  we outline the model used to predict the flux densities 
  for unipolar-inductor-driven radio emission 
  in white-dwarf systems.

\subsection{The unipolar-inductor current circuit} 

\begin{figure}  
\begin{center}   
\vspace*{0.25cm}
\epsfig{file=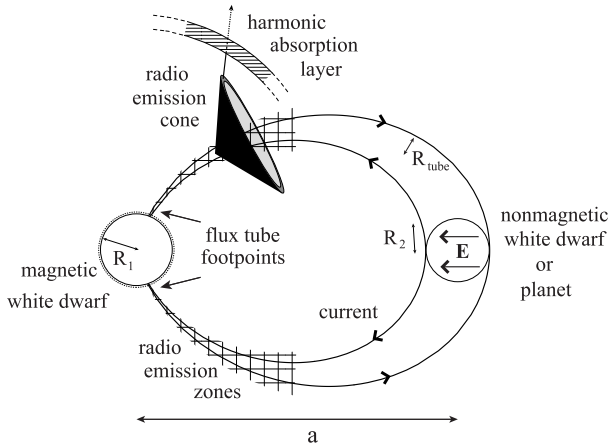,width=8cm}
\end{center} 
\caption{
  The unipolar-inductor model for 
     white-dwarf pairs and white-dwarf planetary systems,
where ${\bf E}$ is the induced electric field across the
nonmagnetic white dwarf or planet. 
     } 
\label{fig:geometry}
\end{figure} 

Figure~\ref{fig:geometry} shows a schematic illustration
  of the unipolar-inductor model used in our calculation.  
For the magnetic-nonmagnetic white-dwarf pair, 
  the system parameters 
  are the magnetic white dwarf's mass $M_{1}$ and magnetic moment $\mu$, 
  the non-magnetic white dwarf's mass $M_{2}$, 
  and the binary period $P$.    
The radii of the magnetic and non-magnetic white dwarfs, 
  $R_{1}$ and $R_{2}$ respectively, 
  are obtained from the Hamada \& Salpeter (1961) mass-radius relation 
  (assuming helium white dwarfs),  
  and the orbital separation $a$ is calculated 
  from the orbital period and the masses of the two white-dwarfs 
  using Kepler's third law.  
For the white-dwarf planetary system, 
  the nonmagnetic white dwarf is replaced by a planet, 
  and we use the same notation for the mass and radius 
  (i.e.\ $M_2$ and $R_2$) for the planet.  
However, the planetary radius is treated as a free parameter 
  instead of a quantity deduced from a mass-radius relation.     

We define an asynchronism parameter $\xi$, 
  which is the ratio of the spin angular speed 
  of the magnetic white dwarf to the orbital angular speed. 
In terms of the orbital parameters and the asynchronism parameter, 
  the e.m.f. across the nonmagnetic white dwarf is 
\begin{equation}
   \Phi = \left(\frac{\mu_{1} R_{2}}{c}\right)
         \left(\frac{2 \pi}{P}\right)^{7/3} (1-\xi)~
         \big[G(M_{1}+M_{2})\big]^{-2/3}  
\end{equation}
  (see Wu et al.\ 2002).   
For white-dwarf pairs with $P \approx 10$~min 
  and $\mu_1 \sim 10^{32}$~G~cm$^3$, 
  a seemingly small departure from synchronous rotation 
  (e.g., $(1-\xi) \sim 0.001$) 
  can produce a sufficiently high e.m.f. to power 
  X-ray emission at the magnetic footpoints 
  with a luminosity a thousand times the solar bolometric luminosity. 

Most of the electrical energy in magnetic-nonmagnetic white-dwarf pairs  
  is dissipated in the flux tube footpoints 
  at the surface of the magnetic white dwarf. 
This is because the cross-sectional area of the flux tube 
  connecting the two white dwarfs at the footpoints 
  is much smaller than the cross-sectional area at the nonmagnetic white dwarf. 
White-dwarf planetary systems will also dissipate most energy at the footpoints, 
  provided that the conductivity of the planet 
  is higher than the white-dwarf atmospheric conductivity.  
This condition is satisfied for a planetary body 
  composed entirely of an Earth-like core,  
  after the outer planetary layers have evaporated away at an earlier stage 
  in the system's evolution.
In the case where a mantle layer is still present, 
  the bulk of the electrical power would heat the planet
  if the mantle conductivity is lower than 
  the white-dwarf atmospheric conductivity (Li et al.\ 1998).  
We consider only the case with an Earth-like planetary core in this study.

The effective resistance of the magnetic white dwarf 
  at the footpoints of the flux tube is 
\begin{equation}
 {\cal R}_{1} \approx 
    \frac{1}{2 \sigma_{1} R_{2}} \left(\frac{a}{R_{1}}\right)^{3/2}  \ ,   
\end{equation}  
  (see Wu et al. 2002)
  where $\sigma_{1}$ is the white-dwarf atmospheric conductivity,
  and $a$ is the orbital separation.
The current flowing in the circuit is  
\begin{equation} 
   I= \frac{\Phi}{{\cal R}_{1}} \ , 
\end{equation} 
   and is confined to a current sheet at the surface of the flux tube. 
The number density of current-carrying electrons $n_{\rm foot}$
  at the footpoint is given by 
\begin{equation}
  n_{\rm foot} = \frac{I}{e v A} \ ,
\end{equation}
  where $e$ is the electron charge,
  $v$ is the average velocity of the current-carrying electrons,
  and $A$ is the area of the arc-shaped footpoint. 
We can approximate $A$ by  
\begin{equation}
   A \approx \pi \, \delta \, R_{\rm tube}(R_{1})^{2} \ ,
\end{equation}
  where $\delta$ is the relative width of the current sheet,
  and $R_{\rm tube}(R_{1})$ is the flux-tube radius at the footpoint.  
Assuming a dipolar white-dwarf magnetic field, 
  the flux-tube radius varies with distance $r$ from the magnetic white dwarf as
\begin{equation}
  R_{\rm tube}(r) \approx R_{2} \left(\frac{r}{a} \right)^{3/2} \ .
\end{equation}
It follows that the number density of current-carrying electrons 
  varies with distance as
\begin{equation}
   n_{\rm c} (r) \approx n_{\rm foot} \left(\frac{R_{1}}{r}\right)^{3} \ .
\label{eq:nvar}
\end{equation}

\begin{table*}
\centering
\begin{minipage}{140mm}
\caption{System parameters: 
   mass $M_{2}$ and radius $R_{2}$ of the nonmagnetic white dwarf (first row) 
   and planet (second and third rows), period $P$,
   asynchronicity parameter $\xi$, induced e.m.f. $\Phi$, 
   footpoint flux-tube radius $R_{\rm tube}(R_{1})$, 
   and number density of current-carrying electrons at footpoint $n_{\rm foot}$.}
\begin{tabular}{@{}ccccccc}
\hline
  $M_2$\ ($M_{\odot}$) & 
 $R_2$\ (cm) & $P$\ (hr) & $\xi$ & $\Phi$\ (statvolt) & 
 $R_{\rm tube}(R_1)$\ (cm) & $n_{\rm foot}$\ (cm$^{-3}$) 
  \footnote{assuming $\delta=0.05$~rad and $v=0.1 c$} \\
 \hline
 $0.5 $ & $9.6 \times 10^{8}$ & 0.14  
    & 0.999 & $4.0 \times 10^{4}$ & $2 \times 10^{7}$ &$1.7 \times 10^{12}$ \\
 $1.2 \times 10^{-6}$ & $3.5 \times 10^{8}$ 
    & 10 & 0 & 960 & $1.4 \times 10^{5}$ & $6.1 \times 10^{12}$ \\
 $1.2 \times 10^{-6}$& $3.5 \times 10^{8}$   
    & 28 & 0 & 88&  $5.0 \times 10^{4}$ & $1.6 \times 10^{12}$ \\
\hline
\end{tabular} 
\end{minipage}
\label{tab:1}
\end{table*}

\subsection{Formation of loss-cone electron distribution}

As the downward-propagating, non-thermal, current-carrying electrons 
  reach the polar footpoints, 
  they are lost to collisions in the dense white-dwarf atmosphere. 
The dissipated electrical power produces the footpoint X-ray emission.   
We assume that collisional losses become significant 
  at a height $0.01 R_{1}$ above the white-dwarf surface, 
  corresponding to the thickness of the white-dwarf atmosphere  
  (e.g.\ Kippenhahn \& Wiegert 1989). 

Assuming adiabatic motion of electrons along the field lines, 
  the electron speed is constant, 
  together with the adiabatic invariant $\sin^{2} \alpha/B$, 
  where the electron pitch angle $\alpha$ is the angle 
  between the velocity ${\bf v}$ and the magnetic field ${\bf B}$.  
Higher up the flux tube (away from the footpoint), 
  the precipitating electrons occupy the range of pitch angles 
  $0 < \alpha < \alpha_{\rm lc}$, 
  with the critical angle $\alpha_{\rm lc}$ given by 
\begin{equation}
  \alpha_{\rm lc} = \sin^{-1}\left[\left(\frac{B}{B_{\rm f}}\right)^{1/2}\right] \, ,    
\end{equation}  
  where $B$ is the local magnetic field strength  
  and $B_{\rm f}$ is the field strength at the footpoint. 
Thus the precipitating electrons map into a decreasing range of pitch angles 
  with increasing distance from the footpoint. 
The other downward-propagating electrons 
  with pitch angles $\alpha>\alpha_{\rm lc}$
  are reflected, before reaching the white-dwarf atmosphere, 
  by a magnetic mirror 
  formed by the converging field lines at the footpoint.
Hence, the distribution of the reflected (upward-propagating) electrons 
  at this position has a deficiency in electrons
  with pitch angles $\alpha< \alpha_{\rm lc}$.  
This is the loss cone. 
The critical (loss-cone opening) angle $\alpha_{\rm lc}$  
  decreases from $\pi/2$~rad, at the footpoint,  
  with increasing distance along the flux tube.

\begin{figure} 
\begin{center}  
\vspace*{0.25cm}
  \epsfig{file=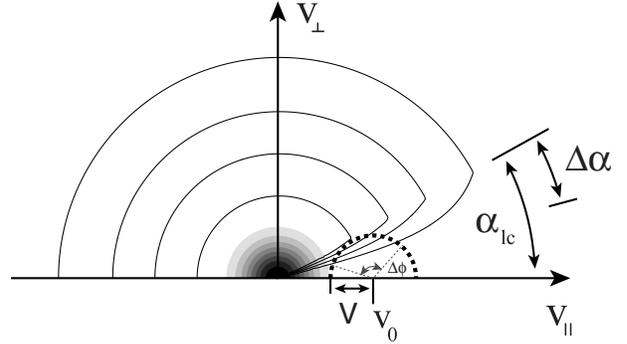,width=8cm}   
\end{center}
\caption{
  The loss-cone distribution is illustrated 
     as contours in $v_{\parallel} - v_{\perp}$ space, where
  $v_{\parallel}$ and $v_{\perp}$ 
     are the components of the velocity ${\bf v}$ 
     parallel and perpendicular to the magnetic field ${\bf B}$.  
  The greyscale contours at low velocities 
     correspond to the thermal background electron distribution, 
     with number density $n_{\rm th}$ and temperature $T_{\rm th} \ll T_{\rm lc}$.  
  A resonance circle (see Appendix \ref{sec:ecme}), 
     with centre $v_{0}$ and radius $V$,
     is illustrated by a dashed line. 
  The angle $\Delta \phi$ corresponds to the segment of the resonance circle 
    which intersects the loss-cone edge.
   }
\label{fig:losscone}
\end{figure}

The loss-cone distribution is modeled here with the Gaussian form:  
\begin{equation}
  f(v,\alpha) = f_{0} (v) \, F(\alpha) \, ,
\label{eq:losscone}
\end{equation}
   with the velocity distribution   
\begin{equation}
  f_{0}(v) = \frac{n_{\rm lc}}{(2 \pi)^{2/3} m_{\rm e}^{3} V_{\rm lc}^{3}}\ 
{\rm exp} \left( - \frac{v^{2}}{2 V_{\rm lc}^{2}} \right) \, ,
\end{equation} 
  and the angular distribution 
\begin{equation}
  F(\alpha) = \left\{ \begin{array}{ll}
     0 & \alpha < \alpha_{\rm lc}- \Delta \alpha \\ 
       1 & \alpha > \alpha_{\rm lc} \\ 
    {\Delta \alpha}^{-1}
   \left[ {\alpha- (\alpha_{\rm lc} - \Delta \alpha)}\right]  
      &  {\rm otherwise} 
\end{array} \right. \, , 
\end{equation}
  where $\Delta \alpha$ is the loss-cone edge width, 
  $n_{\rm lc}$ is the number density of loss-cone electrons,
  and $V_{\rm lc}=(k_{\rm B} T_{\rm lc}/m_{\rm e})^{1/2}$ 
  is the thermal-velocity spread for the loss-cone temperature $T_{\rm lc}$ 
  ($k_{\rm B}$ is Boltzmann's constant). 
The loss-cone distribution 
  is illustrated in Figure~\ref{fig:losscone},  
  as contours in $v_{\parallel}-v_{\perp}$ space 
  (the components of {\bf v} parallel and perpendicular to {\bf B}).    
We note that our calculations assuming an idealized loss-cone distribution
  tend to overestimate the maser growth rate. 
In practice, the maser will operate close to a state of marginal stability 
  due to competition between formation and maser-induced relaxation 
  of the loss-cone instability (Robinson 1991). 
These considerations will be included in a more comprehensive model 
  in a future paper.

We adopt the loss-cone instability in our model 
  based on the following strong evidence  
  for loss-cone-driven maser producing
  Io-controlled Jovian decametric radiation:   
(i) Connerney et al.\ (1993) reported an anti-correlation 
  between the occurrence of Jovian decametric radiation 
  and the presence of an infrared auroral footprint  
  as a function of Jovian longitude.
Radio emission is preferentially 
  emitted at flux-tube footpoint longitudes with higher field strengths,  
  where the magnetic mirror is more efficient. 
The field strength varies as a function of footpoint longitude 
  due to the asymmetries in the 
  tilted and non-centred Jovian magnetic field. 
At these longitudes 
  a higher proportion of electrons are reflected, driving maser emission 
  at the expense of precipitating electrons, 
  which produce the footpoint emission.
(ii) It is well established that the radiation pattern 
  of Jovian decametric radiation 
  is a hollow cone, 
  with opening angle $\la 80^{\circ}$ and  beam width $\la 2^{\circ}$
  (Dulk 1967; Maeda \& Carr 1992; Kaiser et al.\ 2000). 
This is consistent with the theoretical prediction 
  for loss-cone driven masers (Hewitt et al.\ 1981). 
  
We note that an alternative mechanism, 
  where the maser instability is driven by a ``shell'' electron distribution, 
  is also proposed to be relevant to astrophysical contexts  (Ergun et al.\ 2000). 
The shell electron distribution, 
  which drives Earth's auroral kilometric radiation,
  is created by magnetic-field-aligned electric fields in the auroral zones. 
We argue that a shell maser is not applicable 
  to Io-controlled Jovian radio emission  
  because it produces a thin-disc radiation pattern,  
  beamed exactly perpendicular to the magnetic field direction.  
It is unlikely that refraction or scattering  
  can transform thin-disc beams into the observed hollow-cone beams, 
  which have opening angles $\theta \la 80^{\circ}$, 
  and angular widths $\la 2^{\circ}$.    
However, we cannot rule out the operation of the shell maser 
  in white-dwarf systems. 
Nevertheless, the higher efficiency of the shell maser 
  will produce higher flux densities 
  than predicted by the loss-cone maser model adopted in this study,
  thereby improving the prospects for radio observation of white-dwarf systems.

\subsection{Flux density spectra}

Melrose \& Dulk (1982) derived semi-quantitative expressions
  for the loss-cone-driven electron-cyclotron maser growth rate 
  $\Gamma(\omega,\theta)$, 
  emission bandwidth $\Delta \omega/\omega$ 
  and angular spread $\Delta \theta$, 
  as summarized in Appendix \ref{sec:ecme}.
The wave modes generated by the electron-cyclotron maser 
  depend most critically on the ratio $\omega_{\rm p}/\Omega_{\rm e}$ 
  in the source region (Hewitt, Melrose \& R{\"o}nnmark 1981),
  where $\omega_{p}$ is the electron plasma frequency. 
The escaping wave modes 
  (i.e., modes in which waves can escape 
  from the source region  to reach a distant observer)
  favoured by the electron-cyclotron maser 
  are the fundamental and second-harmonic $x$-mode 
  and the fundamental $o$-mode. 
The magnetoionic $x$-mode and $o$-mode 
  have opposite senses of circular polarization.
Higher harmonics have relatively low growth rates 
  and are unlikely to grow to observable levels.  
In Appendix \ref{sec:suppress} 
  we summarize the parameter ranges for these three modes
  in which the electron-cyclotron maser operates.  
  
We estimate the peak brightness temperature in the source region  
  by assuming saturation of the reactive version 
  of the electron-cyclotron maser instability,  
  i.e., where electron phase-bunching effects are significant. 
This assumption is made 
  by analogy to Io-controlled Jovian decametric radio emissions, 
  where the highest flux densities are attained 
  by the subclass referred to as Jovian S-bursts, 
  which exhibit properties consistent with phase-bunching effects 
  (Carr \& Reyes 1999; Willes 2002).    
A reactive growth mechanism saturates at wave amplitudes 
  where electrons become trapped in the field of the growing waves.
The saturation wave amplitude corresponds to 
  where the trapping frequency $\omega_{T}$ 
  becomes comparable with the growth rate, 
  such that electrons are trapped by the wave 
  before they can contribute to wave growth.
The trapping frequency is given by 
\begin{equation}
  \omega_{T} = \left( \frac{e E k}{m_{\rm e}} \right)^{1/2} \, ,
\label{eq:trap}
\end{equation}
  (Melrose 1986, p.70), 
  where $E$ is the wave electric field, 
  and $k$ is the wavenumber, 
  with $k=\omega/c$ for waves with refractive index $n \approx 1$, 
  which is valid for $x$-mode and $o$-mode waves 
  sufficiently above their respective cutoff frequencies.
The peak wave electric field $E_{\max}$ can be obtained 
  by equating the trapping frequency (Eq.~\ref{eq:trap}) 
  with the maser growth rate (Eq.~\ref{eq:growth}), with
\begin{equation}
   E_{\max} = \frac{m_{\rm e} c  \Gamma^{2}}{e \omega } \, . 
\end{equation}
It follows that the peak energy density 
\begin{equation}
  W =  \frac{ E_{\max}^{2}}{4\pi} 
  = \frac{  m_{\rm e}^{2} c^{2}  \Gamma^{4}}
   {4 \pi e^{2} \omega^{2}} \, .
\label{eq:w}
\end{equation} 
The relation between brightness temperature $T_{\rm b}$ 
  and energy density $W$ in the source is  
\begin{equation}
   W= k_{\rm B} T_{\rm b} \left(\frac{\omega}{2 \pi c} \right)^{3} 
      \left( \frac{\Delta \omega}{\omega} \right) \Delta \Omega \,  
\end{equation} 
  (Melrose \& Dulk 1982), 
  where $\Delta \omega/ \omega$ is the bandwidth 
  and $\Delta \Omega$ is the solid angle filled by the radiation, 
  with $\Delta \Omega = 2 \pi \Delta \cos \theta$. 
Using the derived expressions for the bandwidth (Eq.~\ref{eq:bandwidth}) 
  and angular width (Eq.~\ref{eq:deltatheta}) of the loss-cone-driven maser, 
  the brightness temperature  
\begin{equation}
  T_{\rm b} = \frac{4 \pi^{2} c^{6} W}
    {k_{\rm B} \omega^{3} v_{0}^{3} \, \Delta \alpha^{2} \,
\cos^{4} \alpha_{\rm lc}} \, ,
\label{eq:Tb}
\end{equation} 
where $v_{0}$ is related to the emission angle $\theta$ in
Equation (\ref{eq:v0}).
The flux density $S$ is related to the brightness temperature by 
\begin{equation}
  S = \left( \frac{\omega}{2 \pi c} \right)^{2} k_{\rm B} 
     \int  d\Omega \ T_{\rm b}
\end{equation}
  (Dulk 1985), where the integral is taken over the projected area of the source 
  for the differential solid angle $d\Omega$. 
Assuming a constant brightness temperature over a source size 
  equal to the flux tube radius $R_{\rm tube}$, 
  this simplifies to
\begin{equation}
  S = \left( \frac{\omega}{2 \pi c} \right)^{2} k_{\rm B} T_{\rm b} 
    \frac{\pi R_{\rm tube}^{2}}{4 d^{2}} \, ,
\end{equation}
  where $d$ is the distance to the source.  

As the maser emission at harmonic number $s$, 
  generated at a frequency $\omega = s \Omega_{\rm e}$,
  propagates away from the magnetic white dwarf, 
  it encounters harmonic absorption bands 
  at $\omega \approx (s+1)~\Omega_{\rm e}$, 
  $(s+2)~\Omega_{\rm e}, \ldots$.  
The first absorption band is illustrated schematically 
  in Figure \ref{fig:geometry}.  
We incorporate harmonic damping effects in our calculations, 
  using the approximate formula for the optical depth 
  of the harmonic absorption bands 
  in Appendix \ref{sec:harmdamp}.

\section{Results and Discussion}
\label{sec:discuss}  
 
We consider three cases 
  to illustrate the electron-cyclotron maser emission in white-dwarf systems. 
The first is a white-dwarf pair, and
  the other two are white-dwarf planetary systems 
  with differing orbital periods.  
The system parameters of these three systems 
  are listed in Table 1.   
In all cases, the mass of the magnetic white dwarf $M_{1}$ 
  is fixed to be 0.7~M$_{\odot}$,  
  its atmospheric conductivity $\sigma_{1}$ is $10^{14} \, {\rm s}^{-1}$,
  and its magnetic moment $\mu$ is $10^{31} \, {\rm G} \, {\rm cm}^{3}$.  
In the white-dwarf planetary systems, 
  the spin of the white dwarf is not synchronized with the planetary orbit. 
We assume here that the spin period of the isolated magnetic white dwarf 
  is significantly greater than the orbital period,
  and without loss of generality we set $\xi =0$. 
The distances to all systems are fixed to be 100~pc.    
For the loss-cone parameters, unless otherwise stated, 
  we assume that $\Delta \alpha = 0.05$, 
  which corresponds to emission angular beam widths 
  on the order of a few degrees 
  (by analogy with the Jupiter-Io system),
  the loss-cone temperature $k_{\rm B} T_{\rm lc} = 1$ keV 
(where 1 eV $= 1.602 \times 10^{-12}$ erg)
  and the footpoint electron-number density $n_{\rm lc} = 10^{8} {\rm cm}^{-3}$.  
The loss-cone number density varies 
  along the flux tube according to Equation~(\ref{eq:nvar}). 
For these parameters we are assuming that 
  only a very small fraction of current electrons ($\la 0.01 \%$) 
  contribute to the loss-cone instability.    
Our flux density estimates are therefore conservative.

\subsection{Magnetic-nonmagnetic white-dwarf pairs}  
\label{sec:wdpairs}

\subsubsection{Flux density spectra}  
\label{fd_spectrum}

\begin{figure*} 
\begin{center}  
\vspace*{0.25cm}
  \epsfig{file=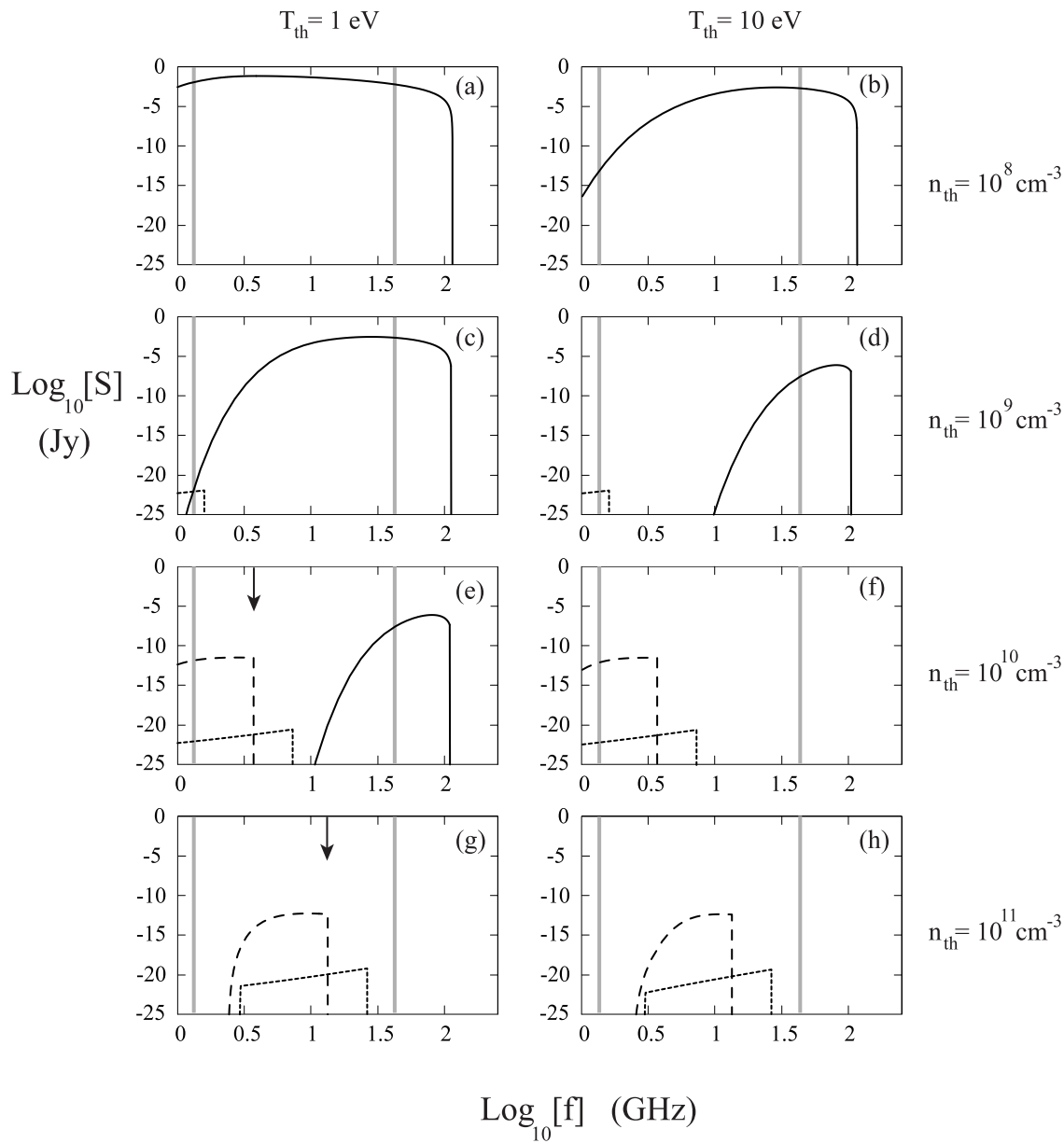,width=15cm}   
\end{center}
\caption{ 
  Peak spectral flux densities for radio emission 
     from magnetic-nonmagnetic white-dwarf pairs 
(maximized over emission angle),
     with binary system parameters in Table~1. 
  The loss-cone parameters are 
     $k_{\rm B}T_{\rm lc} = 1$ keV, 
     $n_{\rm lc} = 10^{8} {\rm cm}^{-3}$, 
     and $\Delta \alpha = 0.05$~rad.  
  Two values are considered for the thermal plasma temperature, with
       $k_{\rm B}T_{\rm th}= 1$ eV (panels a, c, e and g in the left column) 
       and $k_{\rm B}T_{\rm th}= 10$ eV (panels b, d, f and h in the right column); 
     four values are considered for the thermal plasma density, with
       $n_{\rm th}= 10^{8} \, {\rm cm}^{-3}$ (panels a and b),  
       $n_{\rm th}= 10^{9} \, {\rm cm}^{-3}$ (panels c and d), 
       $n_{\rm th}= 10^{10} \, {\rm cm}^{-3}$ (panels e and f), 
       and $n_{\rm th}= 10^{11} \, {\rm cm}^{-3}$ (panels g and h).  
  Spectral density profiles 
     for the fundamental $x$-mode (solid line), 
     the fundamental $o$-mode (dashed line), 
     and the second harmonic $x$-mode (dotted line) are displayed.
  The vertical grey bars denote VLA observation frequencies 
     at 1.465~GHz and 43~GHz.
}
\label{fig:slice}
\end{figure*}  

Figure~\ref{fig:slice} shows the peak spectral flux densities 
  for loss-cone-driven maser emission from a magnetic-nonmagnetic white-dwarf pair  
  over a range of thermal plasma temperatures $T_{\rm th}$ 
  and densities $n_{\rm th}$.   
The binary system parameters are as shown in the first row of Table~1. 
We will discuss the dependence of the peak flux densities 
  on $T_{\rm th}$ and $n_{\rm th}$ immediately below, and subsequently we discuss
   the dependence on the loss-cone parameters (temperature $T_{\rm lc}$, 
  electron number density $n_{\rm lc}$ and width $\Delta \alpha$).    

At low thermal plasma densities 
  ($n_{\rm th}= 10^{8} \, {\rm cm}^{-3}$; panels a and b), 
  the maser operates in the fundamental $x$-mode 
  over the entire displayed frequency range, 
  up to the high-frequency cutoff (above 100 GHz) 
  corresponding to the electron
  cyclotron frequency at the foot of the flux tube.
Reabsorption of fundamental $x$-mode emission 
  at the second harmonic layer (see Appendix \ref{sec:harmdamp}) becomes significant 
  as the thermal temperature is increased. 
Consequently, the flux densities at low frequencies are much lower 
  for $k_{\rm B}T_{\rm th} = 10$~eV than for $k_{\rm B}T_{\rm th} = 1$~eV   
  (compare panels b and a in Figure \ref{fig:slice}).

At intermediate thermal plasma densities
  ($n_{\rm th}= 10^{9} - 10^{10}~{\rm cm}^{-3}$; panels c, d, e and f), 
  fundamental $x$-mode emission is suppressed at low frequencies
  (see Appendix \ref{sec:suppress}).
The fundamental suppression frequency increases 
  with increasing thermal plasma density, 
   as indicated by  the relative positions of the solid arrows
  in Figure~\ref{fig:slice}, panels e and g.
Note that there is a frequency overlap 
  between fundamental and harmonic $x$-mode emission
  because the upper bound to harmonic emission
  is at twice the fundamental $x$-mode suppression frequency.
The abrupt drop in flux density at high frequencies 
  is due to thermal damping
  (see Appendix \ref{sec:ecme}, Equation~(\ref{eq:damp})).  

At high thermal plasma densities 
  ($n_{\rm th}= 10^{11}~{\rm cm}^{-3}$; panels g and h), 
  the fundamental $x$-mode emission generated above the suppression frequency 
  is strongly absorbed. 
Only the fundamental $o$-mode and the second harmonic $x$-mode emission 
  generated below (twice) the suppression frequency
  can escape to an observer. 
No escaping radiation is produced below $\sim 2.5$ GHz, 
  due to suppression of these modes.

\begin{figure*} 
\begin{center}  
\vspace*{0.25cm}
  \epsfig{file=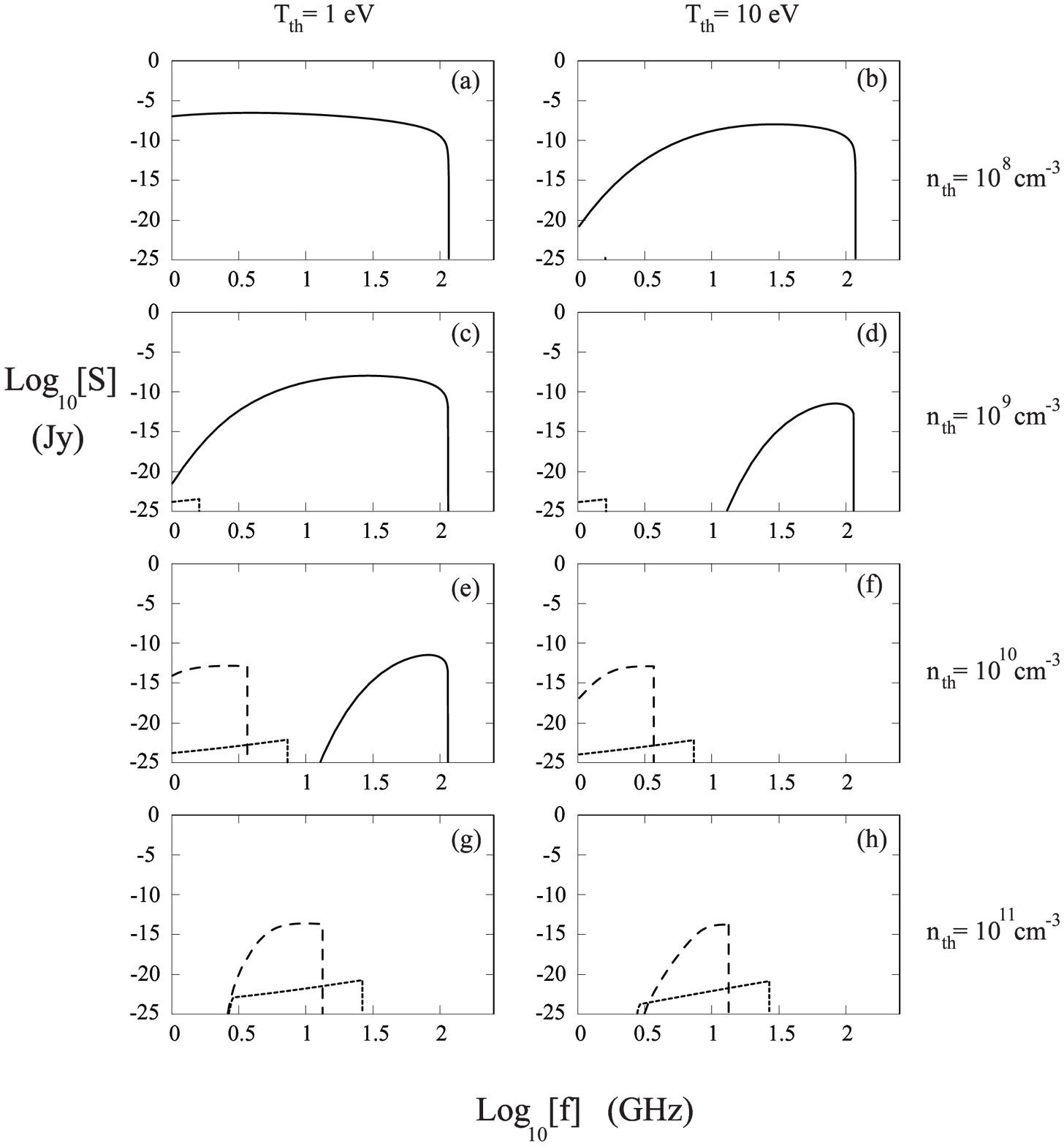,width=15cm}   
\end{center}
\caption{
  Peak spectral flux densities for radio emission 
     from  magnetic-nonmagnetic white-dwarf pairs 
     for same parameters as Figure~\ref{fig:slice}, 
     except with a higher loss-cone plasma temperature: 
     $k_{\rm B}T_{\rm lc} = 10$ keV. 
}
\label{fig:slice2}
\end{figure*} 

The peak flux densities also vary with
  the loss-cone temperature $T_{\rm lc}$.
Figure~\ref{fig:slice2} shows the flux density spectra 
  for the same parameters as Figure~\ref{fig:slice}, 
  except with $k_{\rm B}T_{\rm lc} = 10$~keV. 
The peak flux densities are consistently lower than in Figure~\ref{fig:slice},
  typically by four to five orders of magnitude.
The dependence on the loss-cone electron number density 
  is $S \propto n_{\rm lc}^{4}$ 
  (from Eqs.~(\ref{eq:growth}), (\ref{eq:w}) and (\ref{eq:Tb})).
For example, increasing $n_{\rm lc}$ by a factor of ten, to $10^{8} {\rm cm}^{-3}$, 
  shifts all the flux density curves up by four orders of magnitude
  in Figures \ref{fig:slice} and \ref{fig:slice2}. 
The peak flux densities also decrease with increasing loss-cone width $\Delta \alpha$.
For example, doubling the loss-cone width to $\Delta \alpha = 0.1$ rad 
  lowers the flux density curves in Figures \ref{fig:slice} 
  and \ref{fig:slice2} by an order of magnitude.

\subsubsection{Observability}
\label{sec:obs}

We discuss here the observability of maser emission 
  from white-dwarf pairs by adopting the parameters 
  used in Figure \ref{fig:slice} 
  as an illustrative example.
  We consider two observing frequencies, 1.465~GHz and 43~GHz, 
  which are available for the Very Large Array (VLA) radio telescope. 
We mark these frequencies by grey lines in Figure~\ref{fig:slice}. 
In a recent search for radio emission 
  from Jupiter-like extrasolar planets using the VLA,
  Bastian et al.\ (2000) considered a sensitivity limit 
  of $0.02-0.07$~mJy at 1.465~GHz. 
We assume a sensitivity limit of 0.1~mJy 
  (1 Jy $=10^{-23}~{\rm erg}~{\rm cm}^{-2}~{\rm s}^{-1}~{\rm Hz}^{-1}$) 
  at 1.465~GHz and 43~GHz in our discussion. 
This is a conservative estimate for the sensitivity at 43~GHz 
  because the VLA sensitivity increases with observing frequency.

Maser emission would be detectable at 1.465~GHz 
  for the following ranges of parameters.
The fundamental $x$-mode emission is at least an order of magnitude
  above the detection threshold 
  in panel a of Figure~\ref{fig:slice},
  and hence would be observed over the range of parameters: 
  $n_{\rm lc} \ga 10^{8} \, {\rm cm}^{-3}$, 
  $k_{\rm B} T_{\rm th} \la 1$~eV, 
  and $n_{\rm th} \la 10^{8} \, {\rm cm}^{-3}$.
At higher thermal densities, 
  the fundamental $x$-mode emission is suppressed
  at this frequency (see panel c), 
  and at higher thermal temperatures, 
  the emission is strongly absorbed (see panel b),
  reducing the parameter range for observability to
  $n_{\rm lc} \ga 10^{10} \, {\rm cm}^{-3}$, 
  $1 \la k_{\rm B} T_{\rm th} \la 10$~eV, 
  and $n_{\rm th} \la 10^{8} \, {\rm cm}^{-3}$. 
The fundamental $o$-mode emission is detectable 
  in the parameter range:
  $n_{\rm lc} \ga 10^{10} \, {\rm cm}^{-3}$,
  $k_{\rm B} T_{\rm th} \la 10$~eV,  
  and $n_{\rm th} \approx 10^{10}~{\rm cm}^{-3}$ (panels e and f).

Note that for a loss-cone number density 
  of $n_{\rm lc} = 10^{10}~{\rm cm}^{-3}$, 
  corresponding to $\sim 1\%$ of current-carrying electrons in the loss-cone,
  the flux densities are eight orders of magnitude higher than 
  those in Figures~\ref{fig:slice} and \ref{fig:slice2}
  (with $n_{\rm lc}= 10^{8} \, {\rm cm}^{-3}$). 
This is still a conservative estimate for the fraction of loss-cone electrons, 
  considering that in the Jupiter-Io system
  the observed anti-correlation between footpoint emission and radio emission 
  implies that a more significant fraction of the current-carrying electrons
  contribute to the loss-cone instability.
No radio emission would be detected at 1.465~GHz 
  for $n_{\rm th} \ga 10^{11} \, {\rm cm}^{-3}$, 
  because all free space modes are suppressed 
  at this frequency (panels g and h).

Maser emission would be detectable at 43~GHz 
  over the following ranges of parameters.
The fundamental $x$-mode emission is detectable for  
  $n_{\rm lc} \ga 10^{8}~{\rm cm}^{-3}$,
  $k_{\rm B} T_{\rm th} \la 10$~eV, 
  $n_{\rm th} \la 10^{8}~{\rm cm}^{-3}$ (panels a and b); 
  $n_{\rm lc} \ga 10^{8}~{\rm cm}^{-3}$,
  $k_{\rm B} T_{\rm th} \la 1$~eV, 
  $10^{8} \la n_{\rm th} \la 10^{9}~{\rm cm}^{-3}$ (panel c);
  $n_{\rm lc} \ga 10^{9}~{\rm cm}^{-3}$,
  $1 \la k_{\rm B} T_{\rm th} \la 10$~eV, 
  $10^{8} \la n_{\rm th} \la 10^{9}~{\rm cm}^{-3}$ (panel d); and
  $n_{\rm lc} \ga 10^{9}~{\rm cm}^{-3}$,
  $k_{\rm B} T_{\rm th} \la 1$~eV, 
  $10^{9} \la n_{\rm th} \la 10^{10}~{\rm cm}^{-3}$ (panel e). 

Second-harmonic $x$-mode emission is only relevant 
  in a small region of parameter space 
  (for $n_{\rm th} \approx 10^{11}~{\rm cm}^{-3}$ in panels g and h, 
  in the frequency range $13-26$ GHz. 
A large fraction ($\ga 50\%$) of current-carrying electrons 
  must contribute to the loss cone 
  (corresponding to $n_{\rm lc} \ga 10^{12}~{\rm cm}^{-3}$) 
  in order for second harmonic emission to be detected.
We therefore conclude that second harmonic maser emission 
  only plays a marginal role in radio emission from white dwarf systems.

The predicted radio flux at the Earth from white-dwarf pairs is variable, 
  with zero flux when the emission cone is not directed towards the observer. 
For emission cone widths of several degrees, 
  the duty cycle is on the order of one percent. 
Thus, for the white-dwarf pair parameters assumed above, with $P=500$~s, 
  the radio pulses have durations of $\ga 5$ seconds. 
The interval between pulses is equal to the orbital period $P$ 
  for the extreme viewing geometry 
  where one edge of the emission cone grazes the line-of-sight 
  at one point in the orbit. 
More generally, an emission cone with opening angle $\theta$ 
  will intersect the line-of-sight at two points per orbit, 
  and within each orbital cycle the two pulses 
  are separated by an interval 
  with a value between zero and $P \theta/\pi$, 
  depending on the viewing geometry of the system.

\subsubsection{Unipolar inductor lifetimes} 

An important constraint to the probability of detection for white-dwarf pairs 
  is the lifetime of the unipolar-inductor phase and the binary lifetime.
The duration of the unipolar-inductor phase 
  is related to the rate of synchronization 
  between the magnetic white-dwarf spin and the orbital rotation, 
  due to Lorentz torques associated with the energy dissipation 
  in the current circuit (see Wu et al.\ 2002). 
For the parameters assumed here, this timescale is $\sim 10^{4}$ ~yr. 
The binary lifetime is determined by gravitational radiation power losses 
  and is $\sim 10^{6}$~yr. 
Hence, the probability of detecting close white-dwarf pairs 
  in the unipolar-inductor phase is $\sim 1$\%. 
Despite the low probability of detection, two systems RX~J1914+24 
  (Cropper et al.\ 1998; Ramsay et al.\ 2001, 2002b) 
  and RX~J0806+15 (Israel et al.\ 2002; Ramsay et al.\ 2002a)
  with ultra-short periods ($P <10$~min), 
  are candidate unipolar-inductor white-dwarf pairs 
  and are also potential maser sources.  
If these systems are indeed ultra-short period white-dwarf pairs  
  powered by unipolar induction, 
  their discoveries suggest 
  that there is either an abundant population 
  of such close white-dwarf pairs in the Galaxy,
  or that the unipolar-inductor phase may be longer than 
  the above estimate of $\sim 1$\% of the system lifetime.  
The latter possibility may be due to 
  a close white-dwarf pair entering the unipolar-inductor phase multiple times; 
  for instance, 
  if a (post-unipolar-inductor) synchronous system 
  undergoes intermittent mass exchange 
  that breaks the synchronism and initiates a new unipolar-inductor phase.

\subsection{White-dwarf planetary systems}

White-dwarf planetary systems consisting 
  of a magnetic white dwarf and an Earth-like core 
  with orbital separation of 1~AU ($\sim 2 \times 10^{4} R_{1}$) 
  have orbital periods $\sim 1$~yr.  
The components in these systems are too far apart  
  for the induced e.m.f. across the planetary core 
  to be sufficiently large to accelerate fast (keV) electrons. 
Masers operate more efficiently in more compact systems 
  which are more effective unipolar inductors.  
Here we consider only compact white-dwarf planetary systems   
  with orbital periods $P \approx 10-30$~hr (see Table~1). 
We first discuss the properties of the maser emission 
  from these systems 
  and then other aspects, 
  such as the formation scenario and lifespan of close 
white-dwarf planetary systems. 

\subsubsection{Emission properties} 
\label{WPemission} 

The principle of maser operation 
  in white-dwarf planetary systems 
  is very similar to the white-dwarf pairs discussed above.
The dependence of the emission properties 
  on the loss-cone parameters 
   are qualitatively the same, 
  after rescaling the system parameters (period and masses) accordingly.   
For systems comprising an Earth-sized planetary core 
  orbiting a non-spinning magnetic white dwarf, 
  an orbital period of $\sim 10$~hr 
  corresponds to an orbital separation of $\sim 200~R_{1}$ 
  ($\sim 2~R_{\odot}$).  
As shown in Table 1, 
  the induced e.m.f. in these systems is about 40 times smaller than 
  for the close white-dwarf pairs 
  discussed in \S~\ref{sec:wdpairs}  
  but is comparable to the induced e.m.f. in the Jupiter-Io circuit, 
  which accelerates~keV electrons.  
Moreover, because the size of the planetary core is two orders of magnitude smaller 
  than the size of the nonmagnetic white dwarf, 
  the footpoint radius of the current flux tubes, 
  and hence the maser source size, is greatly reduced 
  in the white-dwarf planetary systems.

\begin{figure*}
\begin{center}  
\vspace*{0.25cm}
  \epsfig{file=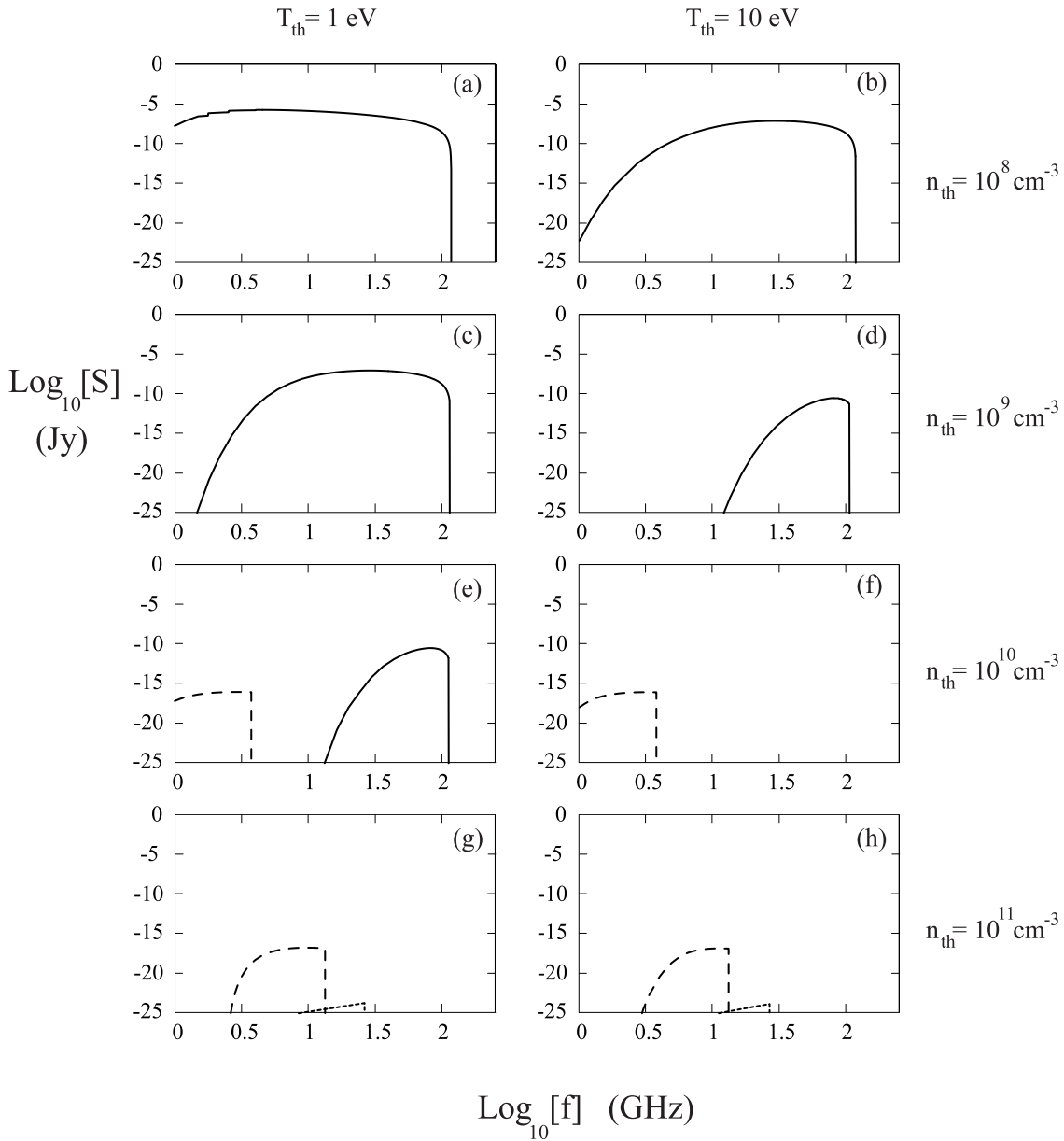,width=15cm}   
\end{center}
\caption{ 
  Peak spectral flux densities for radio emission 
    from an Earth-sized planetary core 
    orbiting a magnetic white dwarf 
(maximized over emission angle)
    for two values of thermal plasma temperature
       ($k_{\rm B} T_{\rm th}= 1$~eV for panels a, c, e, g;
        $k_{\rm B} T_{\rm th}= 10$~eV for panels b, d, f, h), 
    and four values of thermal plasma density
       ($n_{\rm th}= 10^{8} \, {\rm cm}^{-3}$ for panels a and b;
        $n_{\rm th}= 10^{9} \, {\rm cm}^{-3}$ for panels c and d;
        $n_{\rm th}= 10^{10} \, {\rm cm}^{-3}$ for panels e and f;
        $n_{\rm th}= 10^{11} \, {\rm cm}^{-3}$ for panels g and h). 
  The orbital parameters are shown in second row in Table 1. 
  The loss-cone parameters are 
     $k_{\rm B} T_{\rm lc} = 1$ keV, 
     $n_{\rm lc} = 10^{8} {\rm cm}^{-3}$  
     and $\Delta \alpha = 0.05$. 
  Spectral density profiles 
     for fundamental $x$-mode (solid line), 
     fundamental $o$-mode (dashed line), 
     and the second harmonic $x$-mode (dotted line) are displayed.
  }
\label{fig:planet1}
\end{figure*}

We show in Figure~\ref{fig:planet1} the flux density spectra 
  for maser emission from  a white-dwarf planetary system with $P = 10$~hr. 
The predicted flux densities are typically four orders of magnitude lower than 
  for the white-dwarf pair systems in \S~\ref{sec:wdpairs}. 
The parameter ranges for which radio emission is detectable 
  with the VLA are reduced from those quoted in \S~\ref{sec:obs}, 
  with the minimum values of $n_{\rm lc}$ 
  multiplied by a factor of ten 
  (which increases the peak flux densities by four orders of magnitude). 

The peak flux densities decrease as the orbital separation increases. 
For instance, by doubling the orbital separation 
  (corresponding to the system parameters in the third row of Table 1), 
  the induced e.m.f. is an order of magnitude lower, 
  but is comparable to the induced e.m.f. in the Jupiter-Callisto circuit, 
  which generates radio emission of comparable power to the Jupiter-Io circuit
  (Menietti et al.\ 2001). 
The peak flux densities (not shown) are, on average, 
  an order of magnitude lower than in Figure \ref{fig:planet1}; 
  again, the decrease is predominantly 
  due to the smaller source size at the foot of the flux tube.  
  
\subsubsection{Lifetimes and formation scenario} 
\label{aspects}  

Whether or not we can detect maser emission from white dwarf planetary systems 
  depends not only on the emission properties of the systems 
  but also the number of systems 
  and the proportion of the planetary-system lifespan occupied 
  by the unipolar-inductor phase. 
The lifetime of the unipolar-inductor phase 
  is limited by the inward drift of the planet due to Lorentz torques. 
The inward drift speed is estimated to be (Li et al.\ 1998)  
\begin{equation}
   v_{\rm drift} = - \frac{4 \Phi I a^{2}}{G M_{1} M_{2}} \, .
\end{equation}
The characteristic lifetime of the white-dwarf planetary system, 
  $\tau \approx a/v_{\rm drift}$,  
  scales with binary period as $\tau \propto P^{5}$  
  and with magnetic moment as $\tau \propto \mu^{-2}$. 
The 10-hr and 28-hr period systems considered above  
  thus have lifetimes of $\sim 10^{7}$ and $10^{9}$~yr respectively.  
In general, the probability of observing a white dwarf planetary system 
  in the unipolar-inductor phase increases with the binary period $P$. 
This trend is offset by the decrease 
  in the induced e.m.f and maser flux density 
  with increasing $P$.
Increasing the white dwarf magnetic moment 
  increases the induced e.m.f and footpoint dissipation, 
  but simultaneously reduces the unipolar inductor lifetime. 
For example, if we increase the magnetic moment by a factor of 10, 
  to $\mu_{1} = 10^{32} \, {\rm G} \, {\rm cm}^{3}$ for the 28 hour system, 
  the lifetime will decrease from $10^{9}$~yr to $10^{7}$~yr.

There is firm observational evidence 
  for the existence of double degenerate white-dwarf systems 
  and of isolated magnetic white dwarfs.  
Given that the low-mass companion stars 
  in AM Her type binaries and in intermediate polars 
  (see e.g.\ Cropper 1990; Warner 1995)  
  will eventually become degenerate,   
  the existence of magnetic-nonmagnetic white-dwarf pairs  
  is very likely.
However, the existence of white-dwarf planetary systems is far less certain.
We now briefly discuss whether solar-like stellar planetary systems 
  could evolve into the maser-emitting white-dwarf planetary systems 
  investigated theoretically here.   
  
Solar-type stars undergo two major expansions 
  towards the end of their lives 
  --- the first occurs as they reach the Red Giant Branch (RGB), 
  and the second occurs on the Asymptotic Giant Branch (AGB) 
  at the end of the core helium burning. 
The AGB phase is characterized by multiple thermal pulses  
  caused by nuclear instabilities in the helium-burning shell.  
In each thermal pulse,
  the stellar radius expands by approximately $25\%$ 
  and contracts over short timescales (Sackmann et al.\ 1993). 
Following the AGB phase, the star rapidly contracts 
  and eventually becomes a white dwarf. 
  
The expansion phases of a solar-type star have crucial effects 
  on the fate of their orbiting planets. 
There are three likely scenarios:  
(i) Engulfment: 
If the expanding stellar radius exceeds the orbital separation 
  during the RGB phase or the AGB phase (excluding thermal pulses), 
  the planet will evaporate.   
(ii) Temporary engulfment: 
This is when the stellar radius intermittently exceeds
  the orbital separation during the thermal pulses in the AGB phase.
The planet lies within the stellar envelope for a short period
  (between $10^{2}$ and $10^{3}$~yr) 
  and is only partially evaporated 
  (Sackmann et al.\ 1993; Rybicki \& Denis 2001). 
During the temporary engulfment, the dominant force is the ``bow-shock'' drag, 
  which causes the planet to spiral inwards.      
(iii) Otherwise, two competing effects 
  determine the evolution of the planetary orbits. 
The first is that the orbital separation increases  
  due to mass loss from the star in the giant phase (Sackmann et al.\ 1993).
The second effect is that tidal drag forces 
  (through which the orbital angular momentum of the planet 
  is transferred to the stellar envelope) 
  become significant 
  as the stellar radius approaches the orbital separation. 
The tidal interaction acts to decrease the orbital separation,  
  and it dominates over stellar-mass-loss effects 
  when the stellar radius expands beyond a threshold value.    

A prediction for the fate of solar system planets, 
  based on various existing solar evolution models (Rybicki \& Denis 2001),
  is that the planets at a greater distance from the Sun than Mars
  will all migrate out 
  due to stellar-mass-loss effects.  
Mars is beyond reach of the expanded red-giant envelope of the Sun,  
  and the tidal interaction is negligible; 
  so it will not spiral in. 
Mercury and Venus will spiral in and  evaporate.  
The fate of the Earth is less certain, 
  depending on the assumed solar evolutionary model.   
Rybicki \& Denis (2001) estimated the spiral-in distance for the Earth 
  in the temporary engulfment scenario
  to be $(10-70) \, R_{\odot}$ per thermal pulse, 
  and predicted that the Earth can survive several thermal pulses
before being permanently engulfed.
The survival probability increases 
  with the assumed mass-loss rate during the RGB phase in the evolutionary model
  because the number of thermal pulses in the AGB phase is reduced 
  (Sackmann et al.\ 1993).
Clearly, even with uncertainties 
  in the precise evolutionary behaviour of the star, 
  there can still exist a range of initial orbital separations
  from which terrestrial planets migrate to very small final orbital separations 
  at the end of the AGB phase, 
  through a combination of temporary engulfment and tidal interaction effects. 
 
According to this scenario, 
  the parameter space for the formation of close white-dwarf planetary systems is narrow, 
  and we should not anticipate a large fraction 
  of main-sequence planetary systems with terrestrial planets evolving 
  to the compact white-dwarf planetary systems
  that allows efficient unipolar induction to occur, producing strong maser emission. 
The evolutionary scenario discussed above has not taken into account 
  the drift-in effects due to the Lorentz torque (see Li et al.\ 1998).  
Nevertheless, detailed modeling of the planet-star interaction 
  with correct treatments 
  of stellar evolution, hydrodynamics and magnetic interaction 
  are necessary to estimate this fraction.
  


\section{Summary}

We propose that magnetic-nonmagnetic white dwarf pairs 
  with short orbital periods $\la 10$~min, 
  and planetary systems comprising a magnetic white dwarf 
  and an Earth-sized planetary core with orbital periods $\la 30$~hr 
  are strong, periodically variable, and highly polarized radio sources. 
The radio emission is generated by electron cyclotron maser emission
  in a unipolar-inductor current circuit 
  driven by the asynchronicity between the spin of the magnetic white dwarf 
  and the orbital motion of the nonmagnetic white dwarf or planet. 
This model is a direct analogy to Jovian radio emissions 
  generated in conjunction with the Galilean moons.

We apply a simple model to predict the flux densities of the radio emission 
  for the two polarization modes 
  received at the Earth from white-dwarf maser systems, 
  assuming a source distance of 100 pc.
Other model assumptions include a dipolar magnetic field 
  for the magnetic white dwarf, 
  and a constant thermal plasma density in the immediate vicinity of the binary system.
The free parameters in the model are the binary system parameters
  (component masses and radii, orbital distance and period),
  the thermal plasma parameters (density and temperature), and
  loss-cone parameters (density, temperature, and loss-cone width)
  for the electrons which are accelerated in the flux tube current
  circuit connecting the two components.

Our conclusions for magnetized-unmagnetized white dwarf pairs are as follows.  
 (1) White dwarf pairs with orbital periods $\la 10$ min are  
     strong radio sources over a broad range of free parameters in the model.
 (2) The frequency range of radio emission in various wave modes is determined 
     by the primary white dwarf's magnetic moment and the thermal plasma density. 
 (3) The radio emission spectrum can be discontinuous, 
     and the two polarization modes (with opposite senses of circular polarization)
     may be detected at different frequencies.   
 (4) The detection probability is limited by the lifetime of the unipolar inductor phase 
     relative to the binary lifetime.  
 (5) The predicted radio emission is periodic following the orbital motion of the system, 
     with the duty cycle depending on the width of the hollow-cone emission beam. 

Our conclusions for white-dwarf planetary systems are as follows. 
 (1) Earth-sized planetary cores orbiting magnetic white dwarfs  
     with orbital periods $\la 30$~hr are strong radio sources, 
     but over a more limited range of parameters than for white-dwarf pairs. 
 (2) Electrical dissipation in the unipolar inductor circuit 
     causes the planet to gradually drift in towards the white dwarf. 
     White-dwarf planetary systems 
     have a significantly longer unipolar-inductor phase than white-dwarf pairs. 
 (3) The longer orbital period for white-dwarf planetary systems 
     (in comparison to white dwarf pairs) 
     implies longer pulse times: the pulse timescale 
     is about tens of minutes, with pulse separation about tens of hours
     (compared with the pulse timescales of a few seconds 
     and pulse separation of a few minutes for white-dwarf pairs). 
 (4) Only a small fraction of main-sequence-star planetary systems 
     containing terrestrial planets
     evolve to close-orbit white-dwarf planetary systems 
     which produce maser emission. Nevertheless, the large number of
     galactic white dwarfs increases the detection probability. 

\section*{Acknowledgments}

We thank Zdenka Kuncic, Don Melrose, Peter Robinson, and Mike Wheatland 
  for comments and critically reading the manuscript.   
AJW acknowledges support from the Australian Research Council 
  through an ARC Postdoctoral Research Fellowship.

\appendix

\section[]{Electron-cyclotron maser growth rate}
\label{sec:ecme}

In a magnetized plasma, 
  the resonance condition between waves and electrons is 
\begin{equation}
   \omega - \frac{s }{\gamma}\Omega_{\rm e} - k_{\parallel}
     v_{\parallel} = 0 \, ,
\end{equation}
  where $s$ is the harmonic number, 
  $\Omega_{\rm e}$ is the electron cyclotron frequency,
  and $k_{\parallel}$ and $v_{\parallel}$ are the components 
  of the wavevector and velocity parallel to ${\bf B}$.
Assuming the semi-relativistic approximation (see Wu \& Lee 1979),  
  the Lorentz factor simplifies to
\begin{equation}
  \gamma = 1 + \frac{v_{\parallel}^{2}}{2 c^{2}}
     + \frac{v_{\perp}^{2}}{2 c^{2}} \, ,
\end{equation}
  and the resonance condition traces out a circle 
  in the $v_{\parallel}-v_{\perp}$ space (see Fig.~\ref{fig:losscone}), 
  with centre  on the $v_{\parallel}$-axis    
\begin{equation}
   {v_{0}} = \frac{k_{\parallel} c^2}{\omega} \, , 
\label{eq:rccentre}
\end{equation} 
   and radius
\begin{equation}
   {V} =  c ~ \left[ \frac{k_{\parallel}^{2} c^{2}}{\omega^{2}}
      - \frac{2 (\omega - 
      s \Omega_{\rm e})}{s \Omega_{\rm e}} \right]^{1/2} \, .  
\label{eq:rcradius}
\end{equation} 
 
The electron-cyclotron maser growth rate 
  for harmonic $s$  and wave mode $\sigma$ 
  (with $\sigma=-1$ for $x$-mode waves and $\sigma=1$ for $o$-mode waves) 
  is given by    
\begin{eqnarray}
   \Gamma_{{\rm s}, \, \sigma} ({\bf k}) &=& \int {\rm d}^{3} {\bf p} \,
      A_{{\rm s}, \, \sigma} ({\bf p}, {\bf k}) \, \delta(\omega -
      {s \Omega_{\rm e}}/{\gamma} - k_{\parallel} v_{\parallel}) \, \\ \nonumber
    &\times&
      \left( \frac{s \Omega_{\rm e}}{\gamma v_{\perp}}
      \frac{\partial}{\partial p_{\perp}} + k_{\parallel}
      \frac{\partial}{\partial p_{\parallel}}
      \right) f({\bf p}) \,
\label{eq:grate}
\end{eqnarray}
   with
\begin{eqnarray}
    A_{{\rm s}, \, \sigma} ({\bf p}, {\bf k}) 
     &=& \frac{4 \pi^{2} e^{2} c^{2} \beta_{\perp}^{2}}
      {\omega n_{\sigma} (1 + T_{\sigma}^{2}) \, 
      \partial(\omega n_{\sigma})/\partial \omega} \, \\ \nonumber
     &\times&
    \left| \frac{K_{\sigma} \sin \theta + 
    (\cos \theta - n \beta_{\parallel}) T_{\sigma}}{n_{\sigma} \beta_{\perp}
     \sin \theta} \, J_{s} (z) + J'_{s} (z)  \right|^{2}
\end{eqnarray}
  (Melrose 1980; Melrose \& Dulk 1982), 
  where $T_{\sigma}$, $K_{\sigma}$, and $n_{\sigma}$ are
  the axial ratio, longitudinal polarization and refractive index 
  for wave mode $\sigma$, 
  $J_{s}(z)$ is the Bessel function of the first kind (order $s$) 
  and $J'_{s}(z)$ is its derivative.  
The argument of the Bessel function 
  $z= \omega n_{\sigma} \beta_{\perp} \sin \theta/ \Omega_{\rm e}$;  
  and the velocity component 
  $(\beta_{\perp},\beta_{\parallel}) = (v_{\perp},v_{\parallel})/c$.    
The resonance condition appears as 
  the argument in the $\delta$-function in Equation~(\ref{eq:grate}), 
  and the growth rate is evaluated 
  as a line integral over the resonance circle. 
The conditions for driving a perpendicular-driven maser
  (with $\partial f/\partial p_{\perp} > 0$ on the resonance circle) 
  are therefore satisfied for the loss-cone electron distribution 
  modeled by Equation~(\ref{eq:losscone}).

Melrose and Dulk (1982) derived a semi-quantitative approximation
  for the maximum loss-cone maser growth rate, 
  where for a given loss-cone angle $\alpha_{\rm lc}$, 
  and resonance circle centre $v_{0}$, 
  the maximum growth rate corresponds to the resonance circle 
  which touches the inside edge of the loss cone,
  with radius  $V = v_{0}~ \sin \alpha_{\rm lc}$. 
For $x$-mode waves sufficiently above the cutoff frequency of the mode, 
  the refractive index $n \approx 1$, 
  and the resonance circle centre is related to the wave angle $\theta$, with 
\begin{equation}
  v_{0} = c \cos \theta \, .
\label{eq:v0}
\end{equation}
The maximum growth rate is
\begin{equation}
  \Gamma_{{\rm s},\sigma} 
      = \frac{ 2 \pi m_{\rm e}^{2} c^{2} \, A_{{\rm s}, \sigma} \, \  
        f_{0}(v_{0} \cos \alpha_{\rm lc})} {v_{0}} 
     \frac{\Delta \phi(\alpha_{\rm lc}, \Delta \alpha)}{\Delta \alpha} \, ,
\label{eq:growth}
\end{equation}
  where $\Delta \phi$ corresponds to the portion of the resonance circle 
  which passes through the loss-cone edge, 
  as illustrated in Figure \ref{fig:losscone}, with
\begin{equation}
  \Delta \phi(\alpha_{\rm lc}, \Delta \alpha) 
    = \pi - 2 \sin^{-1} 
   \left( \frac{\sin(\alpha_{\rm lc}-\Delta \alpha)}{\sin \alpha_{\rm lc}} \right) \, ,
\end{equation}
  and
\begin{eqnarray}
A_{{\rm s}, \sigma} & = & 
  \frac{\pi \omega_{\rm p}^{2} m_{\rm e} v_{0}^{2} 
   \sin^{2} \alpha_{\rm lc}  \cos^{2} \alpha_{\rm lc} a_{\rm s} 
   (1 + v_{0} \sin^{2} \alpha_{\rm lc} T_{\sigma}/c)^{2}}
   {4 \omega n_{\rm e} (1+T_{\sigma}^{2})} \, \nonumber \\
   &\times&
   \left(\frac{v_{0} \sin \alpha_{\rm lc} \cos \alpha_{\rm lc}}{c} \right)^{2 s - 2} \, 
   \left( 1 - \frac{v_{0}^{2}}{c^{2}} \right)^{s-1} \, ,
\end{eqnarray}
   where $a_{\rm s}= 4 s^{2 s}/[2^{2 s} (s!)^{2}] = 1$ for $s=1, 2$. 
The maximum growth rate occurs at frequency $\omega_{\max, {\rm s}}$,  
   given by 
\begin{equation}
   \omega_{\max, {\rm s}} = s\ \Omega_{\rm e}
   \left( 1+ \frac{v_{0}^{2} \cos^{2} \alpha_{\rm lc}}{2 c^{2}} \right) \, .
\label{eq:maxgrowth}
\end{equation} 

The bandwidth is obtained 
  by considering the range of radii of resonance circles within the loss cone edge, 
  for constant emission angle 
  (corresponding to a  fixed centre of the resonance circle). 
Then, from Equation~(\ref{eq:rcradius}), the bandwidth is
\begin{equation}
\frac{\Delta \omega}{s \Omega_{\rm e}} \approx
    \frac{v_{0}^{2}}{2 c^{2}} \ 
    {\Delta \alpha \,\cos^{2} 
    \alpha_{\rm lc} \sin 2 \alpha_{\rm lc}} \, .
\label{eq:bandwidth}
\end{equation}
This expression generalizes the result 
  obtained by Hewitt et al.\ (1982), for small $\alpha_{\rm lc}$,   
  to all loss-cone opening angles.
The angular width $\Delta \theta$ 
  is estimated from the range of $v_{0}$ 
  (centre of resonance circle) 
  within the loss-cone edge, with
\begin{equation}
 \Delta \cos \theta = \frac{v_{0}}{c}\ 
    {\Delta \alpha\ \cot \alpha_{\rm lc}} \, .
\label{eq:deltatheta}
\end{equation}

The effect of damping by thermal electrons also needs to be considered.
Figure \ref{fig:losscone} shows a
  resonance circle 
  which intersects the $v_{\parallel}$-axis, with $v_{\parallel}= v_{0}-V$,
  at the velocity $v_{*}$
  where the loss-cone distribution
  merges into the thermal background, with
\begin{equation}
  v_{*} \approx V_{\rm th} \sqrt{{\rm ln}\left(\frac{n_{\rm th}}{n_{\rm lc}}\right)} \, ,
\label{eq:vstar}
\end{equation}
   assuming that $V_{\rm th} \ll V_{\rm lc}$ and $n_{\rm lc} \le n_{\rm th}$.
Those resonance circles with
\begin{equation}
  v_{0} - V \le v_{*} \, ,
\end{equation}
  sample the strong damping region 
  associated with the thermal electrons, 
  and the negative contribution to the integrated growth rate 
  in the thermal region typically far outweighs 
  the positive contribution in the loss-cone region.
Hence, from Equations~(\ref{eq:rccentre}), (\ref{eq:rcradius}) and (\ref{eq:vstar}), 
  cyclotron-maser radiation is strongly damped 
  for resonance circle centres less than the critical value
\begin{equation}
  v_{0} \le \frac{V_{\rm th}} {1- \sin \alpha_{\rm lc}}   
     { \sqrt{{\rm ln}\left({\frac {n_{\rm th}}{n_{\rm lc}}}\right)}} \, .
\label{eq:damp}
\end{equation}
In this analysis, 
  we approximate the onset of thermal damping by a step function in
Equation (\ref{eq:damp}).

\section[]{mode suppression}
\label{sec:suppress}

The electron-cyclotron maser can operate in various plasma wave modes; 
  however only waves generated in the magnetoionic $x$-mode and $o$-mode 
  can escape to a distant observer. 
Waves in other modes are trapped in the source region.
For maser emission to be generated in the escaping modes,
  the peak frequency $\omega_{\max, {\rm s}}$ (Equation \ref{eq:maxgrowth})
  must exceed the relevant cutoff frequency, 
  which for the $x$-mode is (Melrose, 1986, p.171)
\begin{equation}
\omega_{\rm x} = \frac{1}{2} \Omega_{\rm e}
   + \frac{1}{2} (\Omega_{\rm e}^{2} + 4 \omega_{\rm p}^{2})^{1/2} \, ,
\end{equation}
  and for the $o$-mode is
\begin{equation}
\omega_{\rm o} =  \omega_{\rm p} \, .
\end{equation}

With increasing plasma density or decreasing magnetic field strength, 
  the ratio $\omega_{\rm p}/\Omega_{\rm e}$  increases such that
$\omega_{\max, {\rm s}}$ lies below the cutoff frequency, 
  and successive harmonics are suppressed.
For $\omega_{\rm p}/\Omega_{\rm e} \la 0.3$, 
  the condition $\omega_{\max, 1} \ge \omega_{\rm x}$ is satisfied.
In this case, growth of the fundamental $x$-mode saturates the maser,  
  because it has a significantly higher growth rate than other modes
  (Melrose, Dulk \& Hewitt 1984). 
For higher values of $\omega_{\rm p}/\Omega_{\rm e}$, 
  fundamental $x$-mode emission is suppressed, 
  allowing wave growth in the fundamental $o$-mode and second harmonic $x$-mode, 
  which have comparable maximum growth rates. 
The growth rates of higher harmonic $x$-mode and $o$-mode waves
  are too low to produce observable radiation. 
Trapped modes may also grow 
  (e.g.,\ $z$-mode, Bernstein modes; Melrose, Dulk \& Hewitt 1984; 
  Willes \& Robinson 1996); 
  however, these can only escape to an observer 
  after conversion to a free-space mode 
  by linear mode conversion or nonlinear wave-wave processes 
  (Melrose 1991; Willes \& Robinson 1996). 
We only consider growth of free-space modes in this paper.
Growth of trapped modes may lower the effective growth rates 
  of the fundamental $o$-mode and second-harmonic $x$-mode 
  assumed in this paper 
  if the maser does not saturate (Melrose, Dulk \& Hewitt 1984). 

Assuming saturation of the maser, 
  only fundamental x-mode emission is generated 
  when the condition $\omega_{{\rm max}, 1} \ge \omega_{\rm x}$ is satisfied. 
Otherwise, fundamental $o$-mode emission is generated in the parameter range 
  $\omega_{\rm o} \le \omega_{{\rm max}, 1} < \omega_{\rm x}$, 
  and harmonic $x$-mode emission is generated 
  for $\omega_{{\rm max}, 1} < \omega_{\rm x} \le  \omega_{{\rm max}, 2}$.
Hence there is a wide range of $\omega_{\rm p}/\Omega_{\rm e}$ 
  over which fundamental $o$-mode and second harmonic $x$-mode 
  are simultaneously generated.
These ranges vary with emission angle $\theta$, 
  through Equations (\ref{eq:v0}) and (\ref{eq:maxgrowth}). 
We note that in a more self-consistent treatment of maser emission, 
  where the maser operates in a state close to marginal stability (Robinson 1991), 
  the fundamental $o$-mode and second-harmonic $x$-mode 
  compete more effectively with the fundamental $x$-mode 
  over a wider range of parameters. 
 
\section[]{Harmonic damping}
\label{sec:harmdamp}

As maser emission, emitted at $\omega \approx s \Omega_{\rm e}$, 
  propagates away from the source 
  the local cyclotron frequency decreases with decreasing magnetic field strength, 
  and the radiation approaches a harmonic absorption layer,  
  where $\omega \approx (s+1) \Omega_{\rm e}$.
Absorption at higher harmonic layers ($s+2$, $s+3$, $\ldots$) may also occur,
  but the absorption strength falls off rapidly with increasing harmonic number.
Assuming a slab geometry, with linearly varying magnetic field strength
  (scale length $L_{\rm B} \approx R_{1}/3$), 
  and constant thermal velocity $V_{\rm th}$,
  the optical depth at the $s$th layer for $x$-mode radiation satisfies
  (Kuncic \& Robinson 1993)
\begin{eqnarray}
\tau_{\sigma, {\rm s}}(\omega,\theta) 
   &=& \frac{\pi L_{\rm B} \omega}{\rm c} \, \frac{s^{2 s} }{2^s s!} \,
\frac{\omega_{\rm p}^{2}}{\omega^{2}} \, 
\left(\frac{V_{\rm th} \sin \theta}{c}\right)^{2 s -2}  \nonumber \\
  &\times&
  \left[\frac{(1+T_{\sigma} \cos \theta)^{2}+T_{\sigma}^{2} V_{\rm th}^{2}/c^{2}}
  {1+T_{\sigma}^{2}} \right]\, .
\end{eqnarray}

\label{lastpage}

\end{document}